# HIGH RESOLUTION MICROSCOPY AND RAMAN SPECTROSCOPIC STUDIES ON THE FRESHEST MUKUNDPURA METEORITE, RAJASTHAN, INDIA: PRESENCE OF NANODIAMONDS.

Chandrasekharam DORNADULA[a*]., Govind, UMMETHALA[b], Tripathi, RAJENDRA PRASAD[C]., Amir Hossein TAVABI[d], Sairam Krishna MALLADI [b], Rafal E Dunin-BORKOWSKI[d] and Dixit, AMBESH[e].

[a] *Izmir Institute of Technology, Turkey (ORCID:* 0000-0003-1534-4319), [b] *Indian Institute of Technology Hyderabad (ORCID:* 0000-0002-2289-2551), [c]*Jai Narayan Vyas University, Jodhpur,* [d]*Ernst Ruska Centre for Microscopy & Spectroscopy with Electrons (*0000-0003-1551-885X*), Forschungszentrum Julich, Germany and* [e] *Indian Institute of Technology, Jodhpur (*0000-0003-2285-0754
 *)*.

## ABSTRACT

Carbonaceous Chondrites have special significance in the stellar evolution and in particular in the evolution of life on earth. The carbonaceous meteorite that fell in Mukundpura village, Jaipur, Rajasthan on 6[th] June 2017 is one such rare CM2 (Carbonaceous Chondrite) carbonaceous meteorite. We carried out high resolution scanning and transmission electron microscopic (TEM) studies on typical thin sections, showing abundant grains of iridium (Ir), pentlandite (NiS), and more interestingly crystalline carbon (C). These crystallite carbon grains resemble nanodiamond like signature in the freshest Mukundpura meteorite. The high-resolution Raman spectroscopic measurements are carried out on the crystalline carbon grains, showing well resolved three distinct peaks with a vibrational mode at 1315 $cm^{-1}$, with the onset of a weak vibrational mode at 1150 $cm^{-1}$, substantiating the observation of nanocrystalline diamond in Mukundpura meteorite. The broad peak centered at 1360 $cm^{-1}$ and 1575 $cm^{-1}$ (as an average), suggest the presence of graphitic carbon as well together with apparent presence of nanocrystalline diamond. The average size of nanocrystalline diamond is ~ 3-5 nm. High iridium content in this meteorite supports the meteoric impact related iridium anomaly in geological stratigraphic boundaries (e.g.Cretaceous-Tertiary boundary) that has caused mass extinction of flora and fauna.

Keywords: Meteorite, Mukundpura, Nanodiamond, Raman Spectroscopy, TEM, Iridium anomaly, pentlandite, Mass extinction.

*CORESSPONDING AUTHOR: Chandrasekharam DORNADULA : dchandra50@gmail.com*

1. **Introduction**

The carbonaceous meteorites (also known as carbonaceous chondrites) contain carbon varying from 1 to 5 % derived from the inter stellar dust particle (known as exogenous organic matter). These carbonaceous meteorites (CM) are found over a wide geographic location (Pearson et al., 2006, Glavin et al., 2018). Thus, these meteorites fall on earth provide exogenous delivery of carbon on earth that provide clue to the origin of life. The frequency of carbonaceous chondrite meteorites falling on Earth is rare (Sephton, 2004). For example, Mukundpura fall took place after 77 years of Erakot fall, in Madhya Pradesh in India. The Mukundpura meteorite is one such rare CM2 Chondrite meteorite (Tripathi et al 2018, Rudraswami et al., 2019, Ray et al., 2019, Kalpana et al., 2021). This meteorite fell in an agricultural farm in Mukundpura village near Jaipur, Rajasthan (Figure 1) India on $6^{th}$ June 2017. The optical photograph of the CM2 (designated as CM2 carbonaceous Chondrite) meteorite and the small pit caused due to the fall are shown in Figure 2. According to the reports published by the Geological Survey of India (GSI, 2017), immediately after the fall, the meteorite was collected by the Geological Survey of India and small splinters that were left behind in the area was collected by one of the authors (Tripathi et al., 2018). According to the Geological Survey of India initial report the meteoritic impact created 45 cm diameter and 15 cm deep pit. The image of pit was also captured by one of the author and the photograph is shown in Figure 1.

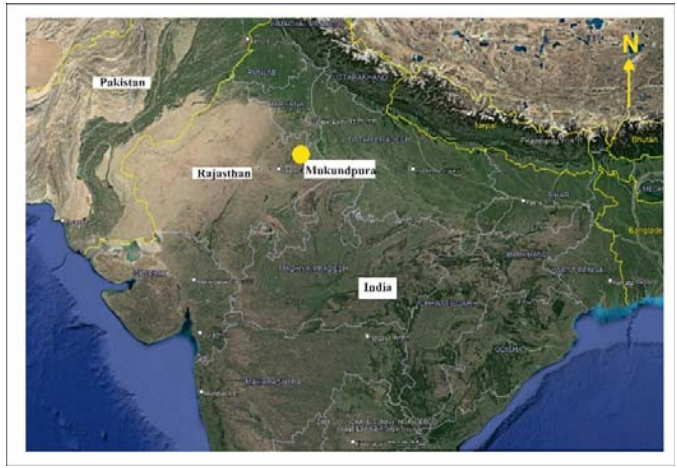

Figure 1- Location of the meteoritic fall in Rajasthan.

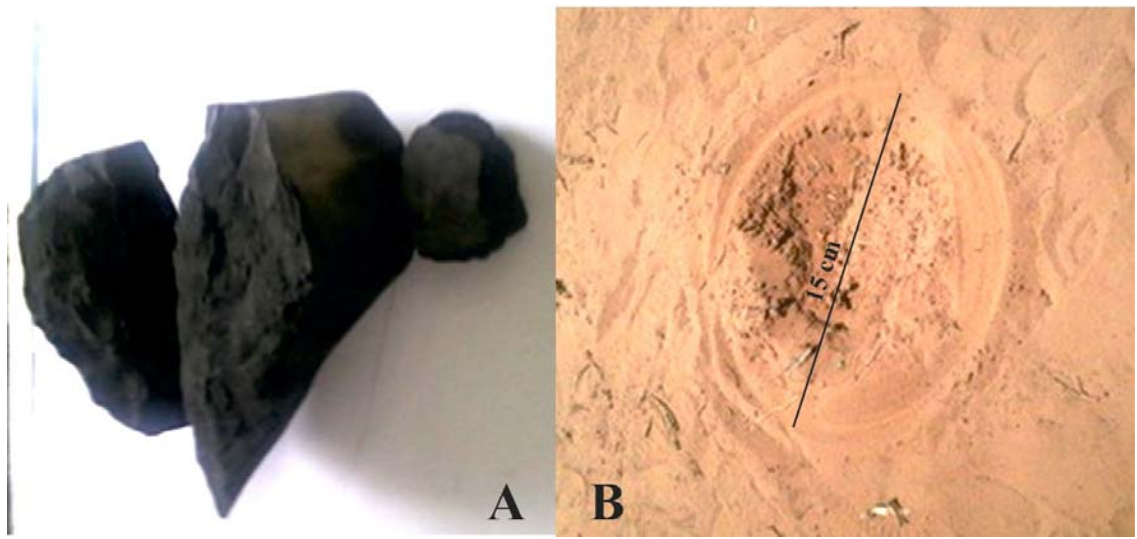

Figure 2- A) The photograph of Mukundpura meteorite and B) Image of pit, created after the meteorite fall (Photo by Tripathi, R.P).

Although the exact weight of the meteorite was not known, GSI (Geological Survey of India) collected about 2.2 kg of the rock fragments. The collected fragments show the presence of fusion crust, formed due to the entry of the mass into the Earth's atmosphere. Petrography, mineralogy (silicate and opaque mineralogy), organic phases (amino acids) present in the same fragment of Mukundpura meteorite, collected by the authors, have been published (Rudraswami et al., 2019, Kaplana et al., 2021). Olivine occur as major chondrules, followed by carbonates and sulphides set in a fine grained matrix (Ray and Shukla, 2018). Presence of water soluble amino-hydroxy and carboxylic acid, amines and their possible enantiomeric excesses have been reported in Mukundpura meteorite (Pizzarello and Yames, 2018). Previous studies have shown the presence of nanocrystalline diamonds both in stony and carbonaceous meteorites (Buseck, and Hual, 1993; Amari et al., 1994; Mostefaoui, et al., 2002; Sephton, 2004; Ott, 2009; Nagashima et al., 2012; Dai et al., 2002*a,b*; Gucsik et al., 2008, Koga, and Naraoka, 2017; Belyanain et al., 2018, Abdu et al., 2018).

The main aim of this work is to report the presence of nanodiamonds in the CM2 meteorite using various techniques such as high-resolution scanning and transmission electron microscopy (HSTEM) and vibrational Raman spectroscopic measurements. This the first time occurrence of nano-diamonds are reported from this meteorite. The understanding of elemental presence is very crucial in the vicinity of nanodiamond like region. The present study also provides the qualitative distribution of different elements present in CM2. The present investigations substantiate the presence of nanocrystalline 3–5 nm crystallite size diamond crystals in Mukundpura meteorite thin sections in addition to the abundance of iridium.

2       **Material and Methods**

Mukundpura meteorite samples were collected by one of the authors (Tripathi R.P) immediately after the fall and details are mentioned in previous reports (Tripathi et al., 2018). Petrographic details of this meteorite have been studied by (Ray and Shukla, 2018). Thick section examination reveal that this meteorite contains 15% of chondrules, 2 % of opaques and the rest constitutes the matrix. Previous studies on this meteorite reveal that the chondrules contains porphyritic olivine, varying in size from 100 to 400 µm (Ray and Shukla, 2018; Murty et al., 1993; Tripathi et al., 2018 ). Pyroxene is rare and if occurs is about 250 μm in size. The mesostasis shows slight alteration. The mesostasis contains isolated olivine grains, several clusters of phyllosilicates. A detailed geochemical analyses of this meteorite can be found in Ray and Shukla (2018) and Rudraswami et al.,(2019). The mesostatis also contains abundant clusters of crystallized phases that can not be identified under the microscope. High content of sulphur rich phases (~10 %) have been identified in the mesostatis ( Ray and Shukla, 2018). The petrographic similarities of Mukundpura meteorite are very similar to the typical Carbonaceous Chondrite of Murchison (Murty et al., 1993). Since this meteorite is not altered and retained its pristinity, it is classified as CM2 type meteorite ( Ray and Shukla, 2018; Murty et al., 1993; Tripathi et al., 2018). Further, the $FeO/SiO_2$ and $S/SiO_2$ ratios match well with the CM2 meteorite of Murchison (Ray and Shukla, 2018; Tripathi et al., 2018). A small chip of 1 cm (Figure 3) has been selected to find possible presence of nano diamonds.

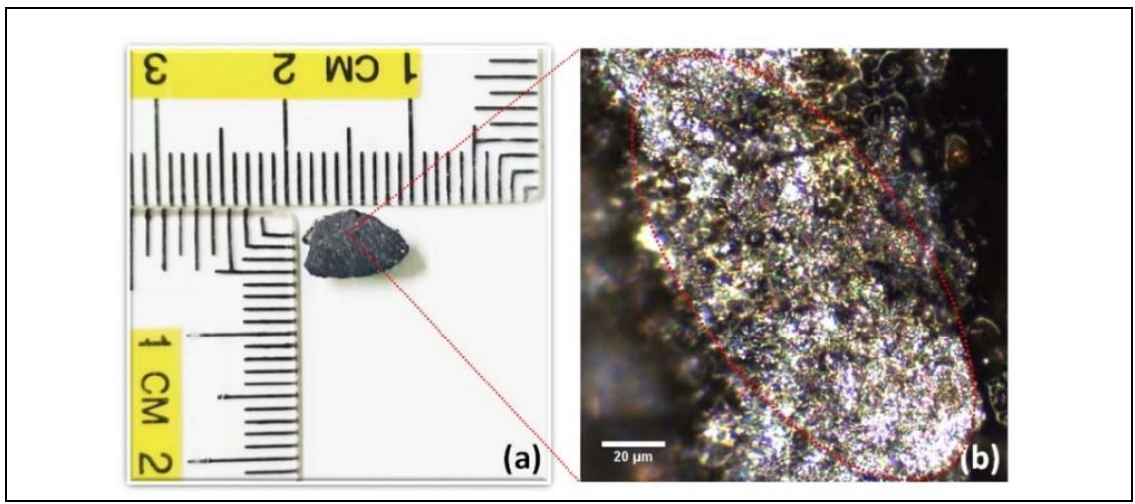

Figure 3- a) Mukundpura meteorite, b) its optical image. The area showing granular texture marked has been selected for further observation.

### 2.1 Sample preparation

A small chip of CM2 is doubly polished to obtain smooth surface for observation under high resolution microscope. The sample has been observed under an optical microscope (Leica),

SEM-EDAX (Scanning Electron Microscope-Energy Dispersive X-Ray analysis) on a JEOL JSM 7800 F, TEMTitan G2 80-200 with ChemiSTEM and Raman spectroscope (WITec alpha 300 R) to explore the presence of nano-diamonds in the area marked on the sample (Figure 3b). This area is selected because of the presence of unusual concentration of crystalline carbon.

### 2.2 SEM images and EDAX analysis

The area marked in the Figure 3b was again observed using a JEOL JSM 7800 F scanning electron microscope. Secondary electron (SE) images were taken from this area and compositional analysis was performed by energy dispersive spectroscopy (EDS) using an Octane Elite EDS detector (EDAX Ametek Instruments) to confirm that it was rich in carbon. The flat polished sample was analysed by performing EDS mapping multiple times using accelerating voltage of 20 kV. The sample is coated with thin gold film. The analysis parameters used are accelerating voltage: 20keV; imaging mode: secondary electron; current beam: 0.17 nA, working Distance: 10 mm; spot size: 6;

### 2.3 Raman Spectroscopy.

Raman spectroscopy was used to identify nan-diamonds in the meteorite studied. The unit is a WITec alpha 300 R system, equipped with an Nd: YAG 532 nm laser. The 2D imaging was performed using the confocal Raman microscope and spectra were collected with the help of a high throughput spectrometer connected to the microscope. The parameters used for Raman mapping are: Model of Raman: alpha 300 R; Points per Line: 100; Lines per Image: 100; Scan Width [μm]: 50.000; Scan Height [μm]: 50.000; Scan Speed [s/Line]: 10.000; Integration Time [s]: 0.1;Excitation Wavelength [nm]: 532.152; Grating: G2: 600 g/mm BLZ=500nm; Center Wavelength [nm]: 595.535; Spectral Center [rel. 1/cm]:1999.999; Objective Name: Zeiss LD EC Epiplan-Neofluar Dic 50x / 0.55; Objective Magnification: 50.0; Laser Spot size: ~ 5 μm.

**2.4. Transmission Electron Microscope imaging**

For TEM imaging Titan G2 Chemi-STEM is used. The specifications of the imaging are: Electron acceleration voltage: 200keV; STEM Detector: Model 3000 HAADF detector. (Fischione); STEM image acquisition: 100s/frame; Mode: HAADF; Spot size: 6;

3  Results :

*3.1. Scanning Electron Microscopic results*

The scanning electron microscopic image of highly polished thin slice of the Mukundpura meteorite sample is shown in Figure 4 and energy dispersive X-ray (EDX) measurement is used to record EDX spectra and elemental mapping for the same sample. The record EDX spectra is shown in Figure 5 and the elemental mapping results are summarized in Figure 4 together with SEM image and marked with respective elements. The presence of Fe-Mg-Ca silicate minerals (olivines and pyroxenes), as seen in thin section, is observed in the SEM images (Figures 4) and EDAX spectrum (Figure 5).

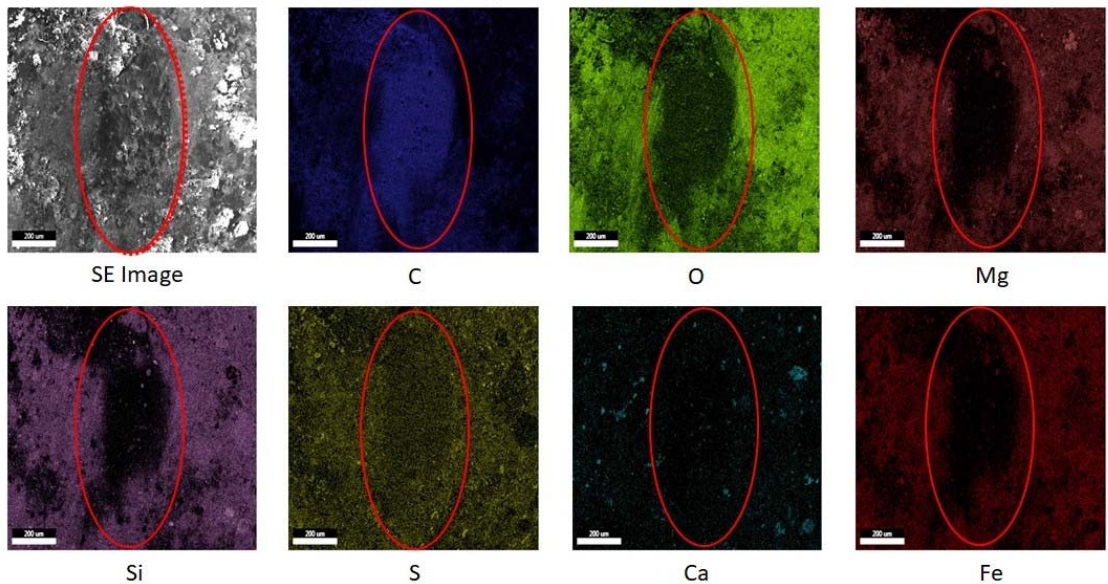

Figure 4- Secondary electron images (of same area as that identified through optical microscope in Figure 2) and corresponding energy dispersive X-ray (EDX) maps, showing the presence of carbon ( C), oxygen (O), magnesium (Mg), silicon (Si), sulphur (S), calcium (Ca) and iron (Fe) various elements the Mukundpura meteorite. Bar scale unit is 200 nm.

The elements are distributed uniformly over the investigated region and the maximum atomic as well as weight fraction is observed for carbon element, as shown in EDX spectra and table, shown in the inset (Figure 5). The presence of carbon (58.9 % atomic or 44.8 wt%) and sulphur (29.7% atomic or 30.1 wt %) in excess of other elements is very clear in the SEM images and EDAX spectrum (Figure 5).

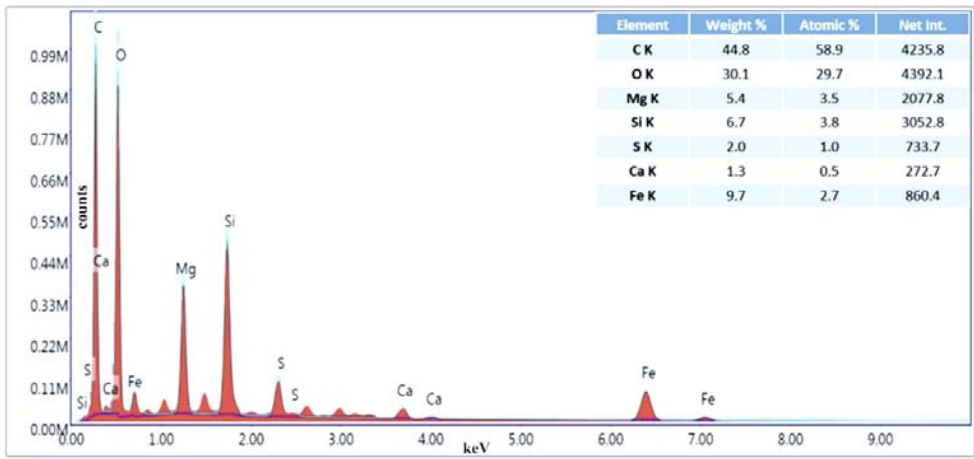

Figure 5- EDX spectra of Mukundpura meteorite, showing the relative presence of various elements, with inset showing relative fraction of different elements with the maximum atomic fraction for carbon.

### 3.2. High Resolution Transmission Electron Microscopic (TEM) Analysis

**3.2.1 Sample preparation**

Small chips of the meteorite sample are carefully selected and cut to a size of about 5 mm to fit into the sample holder of the instrument. The sample is mounted on a slide and the sample is grounded and polished using abrasive paper. Care is taken to avoid pits and groves. The sample is mounted on TEM grid sampler and coated with gold coating gold sputtering machine. Two different samples have been used for SEM and TEM as the samples were analysed in two different laboratories.

3.2.2 TEM analysis

Transmission electron microscope (TEM), Titan G2 with Chemi-STEM was used to obtain high resolution images of the meteorite sample at 200 kV. Further, compositional analysis was performed using EDX (Super-X 4-SDD, windowless EDX detector system). The elemental mappings are shown in Figure 6 and is consistent with the observed SEM-EDX results.

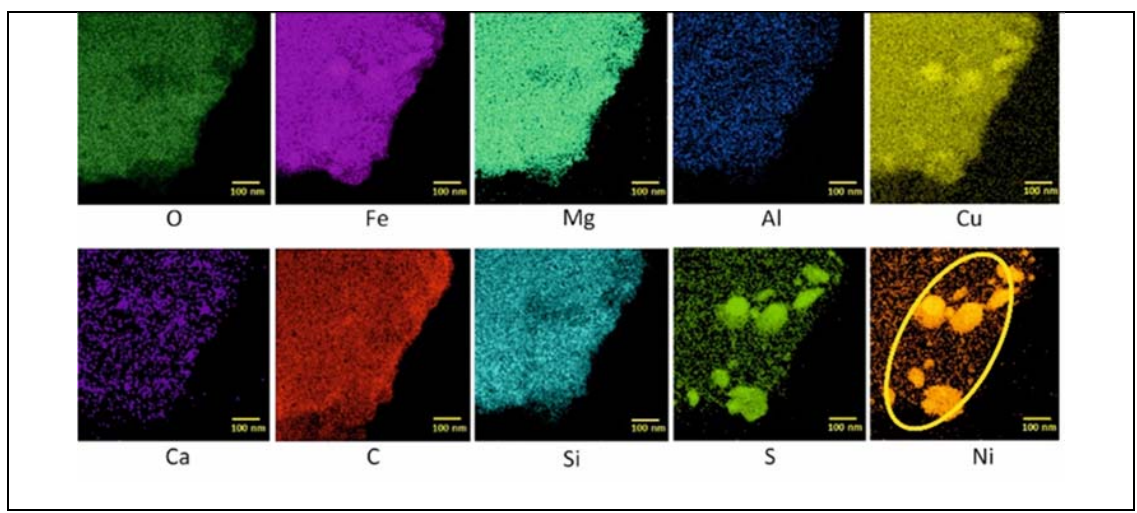

Figure 6- TEM-Elemental mapping for Mukundpura meteorite. Lattice structure of the chondrule in the circled area is shown in Figure 7. Scale bar is 100 nm.

The uniform distribution of elements substantiates the presence of these elements in entire matrix of the Mukundpura sample and are consistent with SEM results, shown in Figure 4. In addition, it was identified that these were nickel rich chondrules, as shown in Figure 6. The chondrules circled in this figure (Figure 6) are subjected to high resolution transmission electron microscopic images, shown in Figure 7. together with the lattice structure of pentlandite of the marked region in high resolution TEM image.

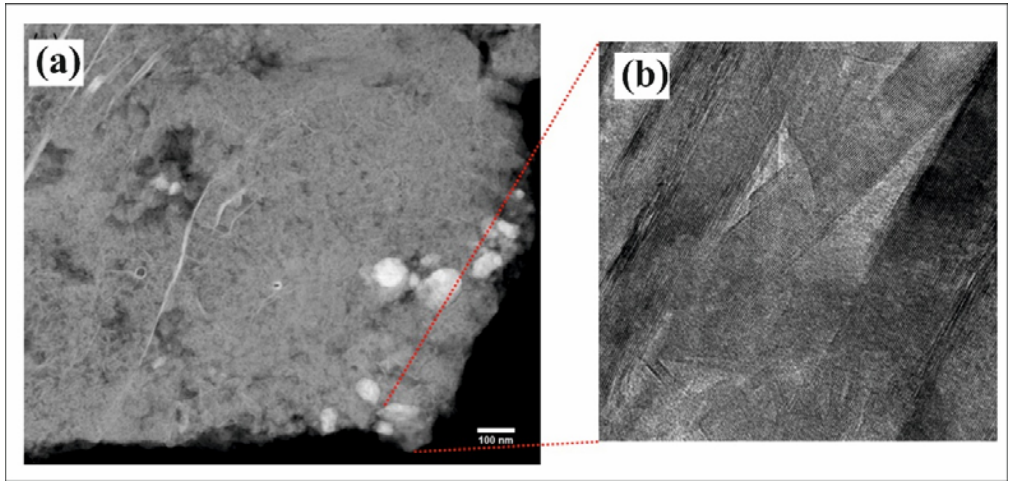

Figure 7- a) High resolution TEM image showing Ni-S (pentlandite) chondrules (white) and b) corresponding lattice structure of pentlandite (100 x 100 nm).

The most striking feature seen in the TEM images is the presence of large Fe-Ni-S chondrules (Figures 7 and 8) that are distributed over the entire sample tested.

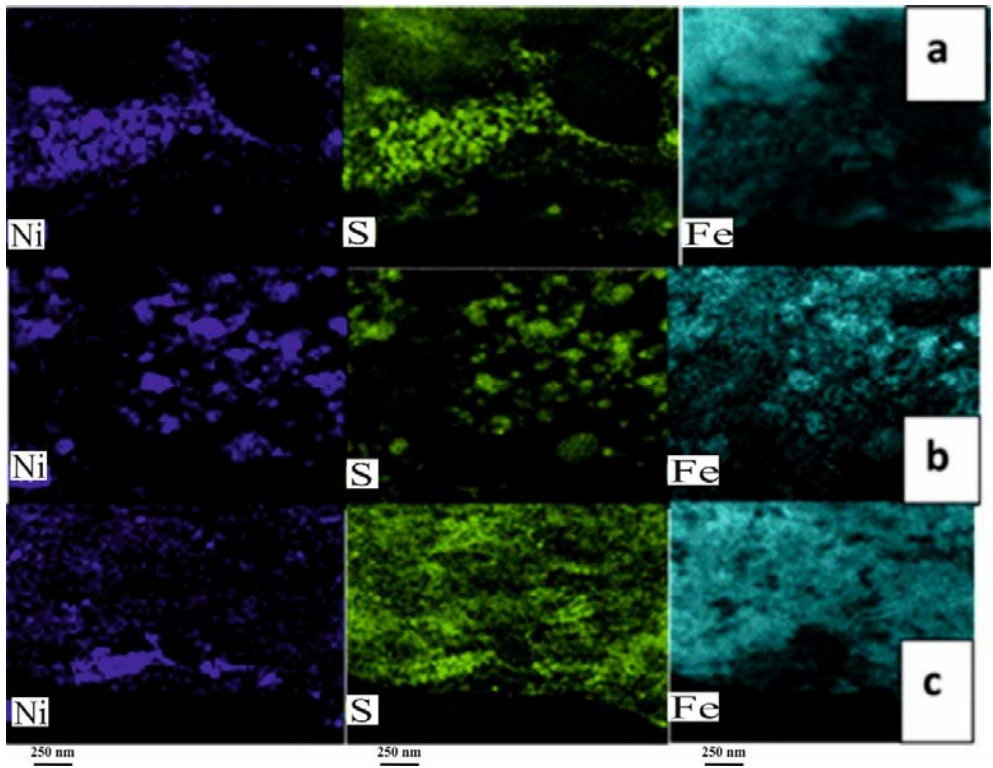

Figure 8- Fe-Ni-S ( $(Fe,Ni)_9 S_8$: Pentlandite) seen under high resolution TEM images, showing Ni, S, and Fe elements presence in Mukundpura meteorite.

These pentlandite chondrules are associated with platinum that occurs as streaks cutting across the matrix or Mg-Fe silicate chondrules (olivine/pyroxene). These are clearly visible in Figure 9.

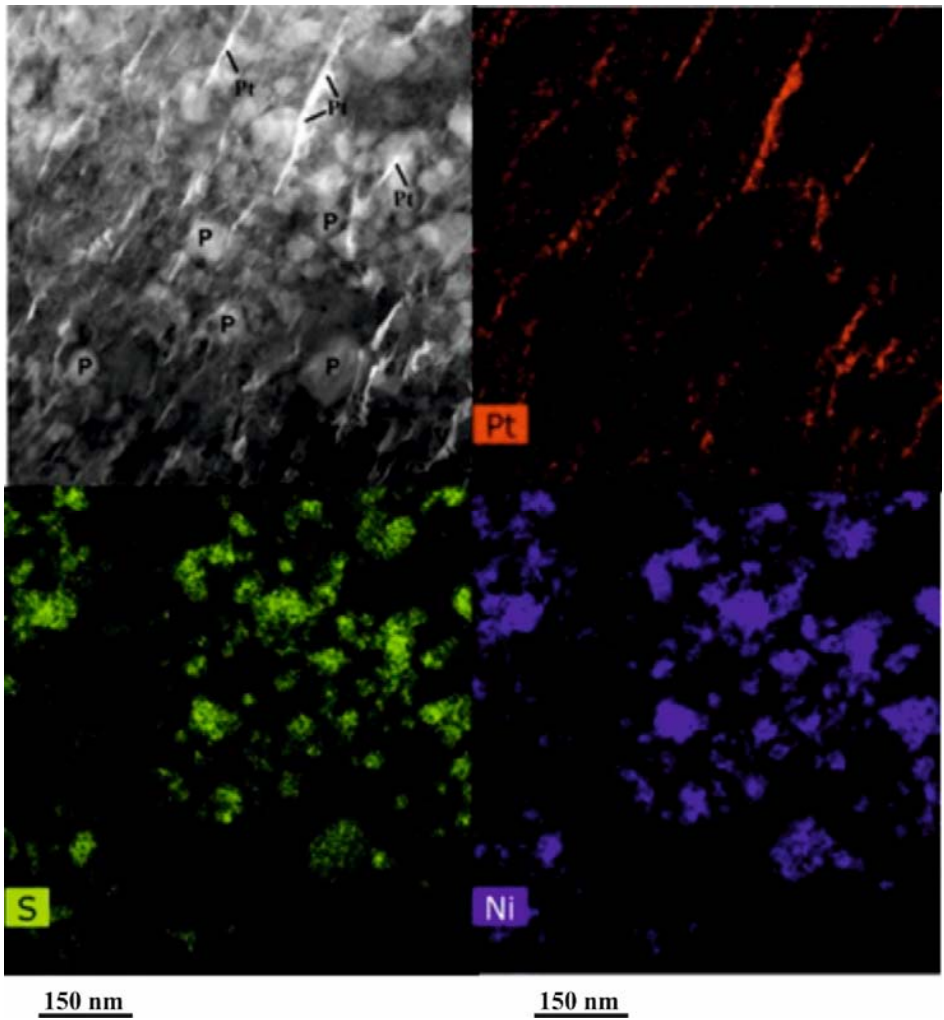

Figure 9- Platinum lamellae (Pt) observed between the granular pentlandites ( P) in Mukundpura meteorite.

### 3.3 Raman Spectroscopy Analysis

The area containing diamonds which was initially identified using optical microscopy (Figure 3) was probed for nano-diamonds. An area scan was performed across the region (140um x 80 um) showing the presence of a cluster of diamond particles. The area map provided an intensity variation of the D and G bands (1360 cm$^{-1}$ and 1590 cm$^{-1}$ peaks) (Figure 10).

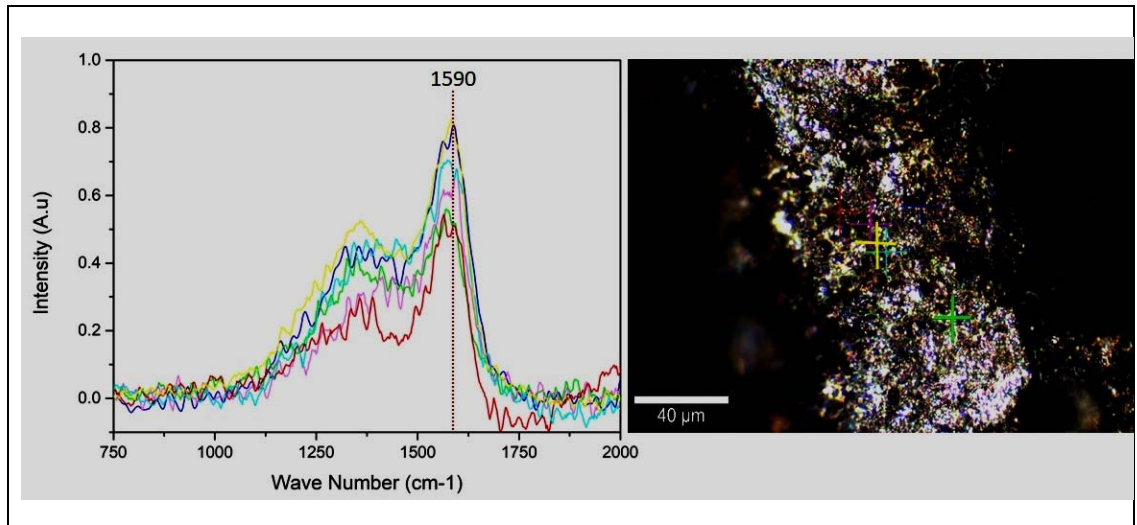

Figure 10- Raman spectra showing diamond peaks. The peaks 1590 cm$^{-1}$ and 1560 cm$^{-1}$ are very similar to the Raman spectra of ultra-nano-crystalline diamonds having an average particle size of ~3–5 nm, i.e., single digit nano-diamonds (Abdu et al., 2018). Each color in the spectra indicates point analysis of the respective color shown in the meteoric section.

## 4 Discussion

High resolution SEM and TEM images of CM2 meteorite reveal the presence of large number of Pentlandite and Platinum chondrules. The amount of crystalline carbon and sulphur is larger than the silicate phases, as noticed in EDX data and elemental mapping. Raman spectroscopy is a non-destructive technique and is used unambiguously to characterize and understand the carbon related materials. The characteristic peak for the diamond single crystal appears at 1332 cm$^{-1}$ with a very high intensity (Murty et al., 1993, Gucsik et al., 2008, Nagashima et al., 2012, Christ et al., 2025). However, the diamonds in such meteorite samples usually exist in small diamond crystalline regions, which may be covered/surrounded with the amorphous or graphitic carbon forming the grain boundaries, and may shed light on the nature of diamond, by producing additional Raman signatures. These additional vibrational modes may assist in understanding the crystallite size and the nature of diamond present in the sample. The single crystal graphite exhibits a single high intensity peak at 1575 cm$^{-1}$ (know as G peak, i.e., crystalline graphite), whereas in other forms of carbon or graphitic materials, an additional peak at 1355 cm$^{-1}$ (known as D peak, i.e., disordered graphite) also appears. The onset of D peak is attributed to the breakdown of selection rules in disordered graphite sample, which is conserved in pure graphite single crystal and thus, the absence of D peak. The intensity of this lower wavenumber peak increases with respect to high wavenumber single crystal graphite peak if the defects (i.e. the randomness or amorphousness increases) in carbon increases together with reducing the graphitic content in the material. The collected Raman spectra, Figure 9, consists of two broad peaks cantered at 1360 cm$^{-1}$ and 1590 cm$^{-1}$, suggesting the presence of both disordered carbon in Mukundpura meteorite. However, these broad peaks consist of several small peaks, as shown in Figure 9 and the deconvolution of broad 1360 cm$^{-1}$ peak, results in three vibrational peaks, cantered at 1320, 1360, 1390 cm$^{-1}$, marked with vertical lines for easy identification, (Figure 9), whereas the peak at 1585 cm$^{-1}$ can be deconvoluted into two clear peaks cantered at 1562, and 1593 cm$^{-1}$, marked with vertical lines (Figure 10). The crystalline and microcrystalline diamond will exhibit a clear distinguishable 1332 cm$^{-1}$ peak with relatively high intensity. However, in microcrystalline diamond may exhibit the presence of disordered or amorphous carbon vibrational modes as well (Ferrari et al., 2012). When the size of diamond crystallites reduces to nanometre range, the identification of corresponding vibrational mode becomes difficult. It is attributed to the breaking of long range ordering, present in single crystal diamond and thus, breaking the selections rules, giving rise to additional vibrational modes in the disordered samples. It may result in shifting of the main 1332 cm$^{-1}$ peak or its asymmetrical broadening together with the onset of new vibrational peaks because of the disordered induced symmetry breaking. The main crystalline 1332 cm$^{-1}$ peak is shifted to lower wavenumber at 1320 cm$^{-1}$ in Mukundpura meteorite sample and is consistent with principle of observing shift and additionally, the peak is also showing asymmetry in its nature. This is a fingerprint evidence of the presence of nanodiamond crystallites in Mukundpura meteorite. Further, the observed G mode cantered at 1585 cm$^{-1}$ and D band at 1560 cm$^{-1}$ together with small intensity ratio of D to G band, i.e., I(D)/I(G) (Gucsik et al., 2008, Nagashima et al., 2012, Ferrari and Robertson,2004) substantiate the presence of crystalline carbon in graphitic form, consistent with SEM and TEM observations. In addition to these diamond, G, and D bands a small but clear vibrational mode is observed at around 1150 cm$^{-1}$, which is not present in the single crystalline or microcrystalline diamond Raman spectra and is considered as the direct evidence of nanocrystalline diamond crystallites. These results provide evidence for nanocrystalline diamond in Mukundpura meteorite sample.

Comparing the present nano-diamond size with nano-diamonds reported from Kapoeta meteorite (Abdu et al., 2018) the nano-diamonds in Mukundpura meteorite appear to be less than (~400 μm in size). Absence of 1450 cm$^{-1}$ vibrational mode and broadening and downward shift of the Raman spectra (Ferrari and Robertson, 2004, Christ et al., 2025, Salek et al., 2025) exclude the possibility that the nano-diamonds in Mukundpura meteorite being formed due to the shock/impact during and after the fall. Further, absence of any terrestrial weathering and the lack of planar deformation features in olivine and absence of major shock induced fabric suggest that the nano-diamonds are inherent to the Mukundpura meteorite and formed during the formation of the chondrite in interstellar space (Ray and Shukla, 2018).

The most interesting feature of this meteorite is the presence of large grains of Iridium (Figure 11), detected using high resolution TEM measurements. Presence of large iridium anomaly between the stratigraphic boundary (Cretaceous-Tertiary) indicate large meteoritic impact during this period that has caused extinction of certain fauna and flora on Earth (Bonte et al., 1984, Elliot et al., 1994, Miller et al., 2010).

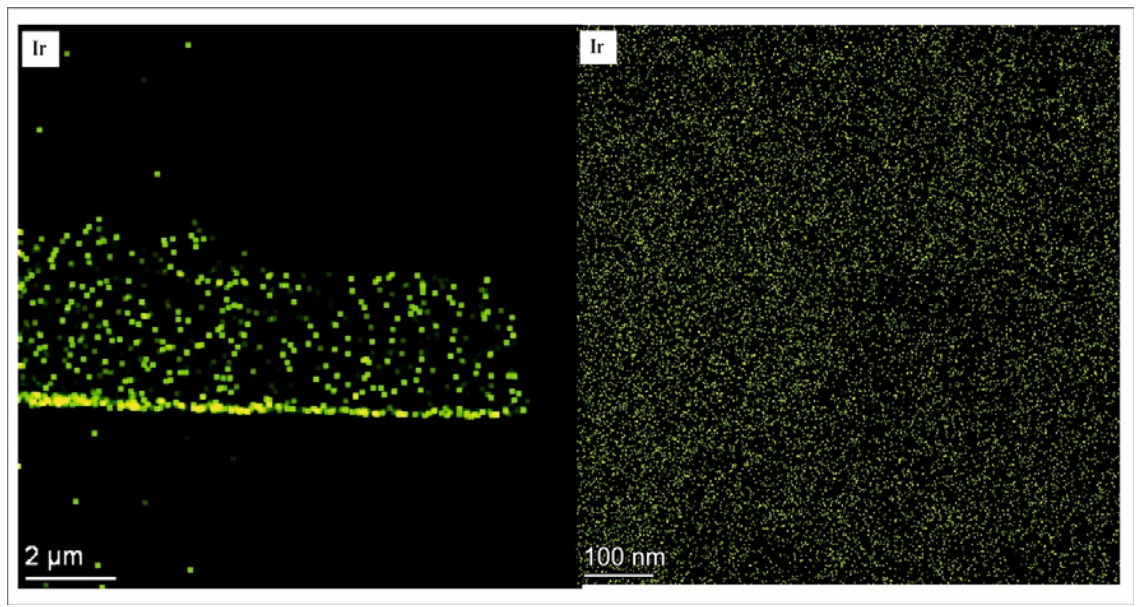

Figure 11. High resolution images of Iridium grains distribution in Mukundpura meteorite.

## 5    Conclusion

High resolution SEM, TEM and Raman spectroscopic investigation on the Mukundpura meteorite reveal the presence of crystalline carbon together with the presence of nano-diamonds. The high-resolution Raman spectroscopic measurements showed the evidences of nanocrystalline size diamond in Mukundpura meteorite. The intensity of these modes is relatively weak due to their nanocrystallite size and presence in the complex material matrix in dilute concentration in nanograins. The average size of nanocrystalline diamond is ~ 3-5 nm. The pentlandite and iridium are found in abundance in CM2. Quantitative estimation of elemental content in the meteorite is in progress using Atom-probe tomography.

**Acknowledgements**

The authors thank the director, IITH, Ernst Ruska Centre for Microscopy & Spectroscopy with Electrons, Forschungszentrum Julich, Germany for extending laboratory facilities to carry-out this work.

Conflict of interest: None